\documentclass[manuscript]{aastex6}

\fullcollaborationName{}
\begin{document}

\title{The Second Galactic Center Black Hole?;  A Possible Detection of Ionized Gas Orbiting around an IMBH embedded in the Galactic Center IRS13E complex}

\author{Masato Tsuboi\altaffilmark{1} and Yoshimi Kitamura}
\affil{Institute of Space and Astronautical Science(ISAS), Japan Aerospace Exploration Agency,\\
3-1-1 Yoshinodai, Chuo-ku, Sagamihara, Kanagawa 252-5210, Japan}
\author{Takahiro Tsutsumi}
\affil{National Radio Astronomy Observatory,  Socorro, NM 87801-0387, USA}
\author{Kenta Uehara}
\affil{Department of Astronomy, The University of Tokyo, Bunkyo, Tokyo 113-0033, Japan}
\author{Makoto Miyoshi}
\affil{National Astronomical Observatory of Japan, Mitaka, Tokyo 181-8588, Japan}
\author{Ryosuke Miyawaki}
\affil{College of Arts and Sciences, J.F. Oberlin University, Machida, Tokyo 194-0294, Japan}
\and
\author{Atsushi Miyazaki}
\affil{Japan Space Forum, Kanda-surugadai, Chiyoda-ku,Tokyo,101-0062, Japan}
\altaffiltext{1}{tsuboi@vsop.isas.jaxa.jp}

\begin{abstract}
The Galactic Center is the nuclear region of the nearest spiral galaxy, Milky Way, and contains the supermassive black hole with $M\sim4\times10^6$ M$_\sun$, Sagittarius A$^\ast$ (Sgr A$^\ast$).  One of basic questions about the Galactic Center is whether Sgr A$^\ast$ alone exists as a ``massive" black hole in the region or not. The IRS13E complex is a very intriguing IR object which contains a large dark mass comparable to the mass of an intermediate mass black hole (IMBH) from the proper motions of the main member stars. However, the existence of the IMBH remains controversial. There are some objections to accepting the existence of the IMBH. In this study, we detected ionized gas with a very large velocity width ($\Delta v_{\mathrm{FWZI}} \sim 650$ km s$^{-1}$) and a very compact size ($r\sim400$ AU) in the complex using ALMA. We also found an extended component connecting with the compact ionized gas. The properties suggest that this would be an ionized gas flow on the Keplerian orbit with high eccentricity.
The enclosed mass is estimated to be  $10^{4}$ M$_\sun$ by the analysis of the orbit. The mass does not conflict with the upper limit mass of the IMBH around Sgr A$^\ast$ which is derived by the long-term astrometry with VLBA.  In addition, the object probably has an X-ray counterpart.  Consequently, a very fascinated possibility is that the detected ionized gas is rotating around an IMBH embedded in the IRS13E complex.
\end{abstract}

\received{}
\accepted{}

\section{Introduction}
The Galactic Center is the nuclear region of the nearest spiral galaxy, Milky Way, and harbors the Galactic Center black hole,  Sagittarius A$^\ast$ (Sgr A$^\ast$)\citep[$M\sim4\times10^6$ M$_\sun$; e.g.][]{Ghez, Gillessen, Schodel2009, Boehle}. 
Its environment is unique in the galaxy,  in which the region contains several peculiar objects.
One of basic questions about the Galactic Center is whether  Sgr A$^\ast$ alone exists as a ``massive" black hole ($\gtrsim10^4 $M$_\odot$) in the region or not.

The IRS13E complex is a very intriguing IR object in the vicinity of  Sgr A$^\ast$, which has been identified  in the early days of the Galactic Center observations. 
It is located approximately 3.5$\arcsec$ southwest of Sgr A$^\ast$ in projection ($r=0.13$ pc) \citep[e.g.][]{Genzel1996, Maillard}.
The center position of the complex corresponds to the west edge of the Minicavity, which is a hook-like substructure of the Galactic Center Minispiral (GCMS) \citep[ e.g.][]{Lacy1980, Ekers1983, LO1983}.
The IR observations at the time suggested that the IRS 13E complex contains several  massive stars including  Wolf-Rayet (WR) and O stars in a diameter of about 0.5$\arcsec$, which have the common direction and similar amplitude of the proper motions (westward proper motion with $V_{mean}\sim 280$ km$^{ -1}$ )\citep[e.g.][]{Maillard}. 
This fact indicates that the main members of the complex are physically bound although the complex should be disrupted by the strong tidal force of Sgr A$^\ast$ \citep[e.g.][]{Gerhard}. 
One possible speculation was that a dark mass like an intermediate mass black hole (IMBH) in the complex may prevent its tidal disruption \citep[e.g.][]{Maillard}. The mass of the IMBH was estimated to be $10^{4}$ M$_\sun$ from the proper motions of the member stars \citep[e.g.][]{Maillard, Schodel2005, Paumard}.
However,  some objections to the existence of the IMBH emerged from new observations  and enviromental evidence \citep[e.g.][]{Schodel2005,  Fritz2010}.
Especially, a recent IR spectroscopic observation indicates that most stars previously detected in the central area of the IRS 13E complex are not massive stars but ionized gas blobs although a few massive stars are certainly identified in the outer area \citep{Fritz2010}.

We consider that the existence of the IMBH in the IRS13E complex is still an open question.
If the IMBH exists,  Atacama Large Millimeter/Submillimetr Array (ALMA) can detect the ionized gas accreting onto the IMBH in the IRS13E complex. The accreting ionized gas is expected to have a very large velocity width and a very compact size.
We searched such ionized gas with ALMA in order to prove the IMBH hypothesis. 
Throughout this paper, we adopt 8.0 kpc as the distance to the Galactic center \citep[e.g.][]{Ghez, Gillessen, Schodel2009, Boehle}; $1\arcsec$ corresponds to 0.04 pc at the distance.

\section{Observation and Data Reduction}
We performed the following observation and analysis. The calibration and imaging of both the data were done by Common Astronomy Software Applications (CASA) \citep{McMullin}. 
\subsection{340 GHz continuum}
We observed  the continuum emission of  Sgr A$^\ast$ and the GCMS at 340 GHz as an ALMA Cy.3 program (2015.1.01080.S. PI M.Tsuboi).
The observations were performed in three days (23 Apr. 2016, 30/31 Aug. 2016, and 08 Sep. 2016). The observation in April was for detection of extended emission.
The field of view  in FWHM (FOV) is 18$\arcsec$, which is centered at Sgr A$^\ast$: $\alpha_{\rm ICRS}$ = $17^{\rm h}45^{\rm m}40\fs04$ and $\delta_{\rm ICRS}$= $-29^{\circ}00'28\farcs20$.
J1717-3342 was used as a phase calibrator. The flux density scale was determined using Titan and J1733-1304. 
We used the self-calibration method in CASA to improve the dynamic range of the map.
The angular resolutions using ``natural weighting" and ``uniform weighting" are $0\farcs14 \times 0\farcs13, PA=82.5^\circ$ and $0\farcs101 \times 0\farcs090, PA=3.8^\circ$, respectively.  The sensitivities are $1\sigma=0.10$ and $0.18$ mJy beam$^{-1}$ in the emission-free areas, respectively.
The dynamic ranges are $\sim28000$ and $\sim16000$, respectively. They are several times worse than the expected values by the ALMA sensitivity calculator. We will present  the improvement of the analysis and the full results in another paper  \citep{Tsuboi2017b}.

\subsection{H30$\alpha$ recombination line}
We analyzed the DDT observation toward the Galactic Center with ALMA, which was released in the autumn of 2016 (Project code: 2015.A.00021.S). The data set contains two-days observations (12/13 July 2016 and 18/19 July 2016) of the Galactic Center in the H30$\alpha$ recombination line ($\nu_{rest}$= 231.9 GHz).
The FOV is 25$\arcsec$, which is centered at Sgr A$^\ast$.  Since the original observations aimed for the measurement of the time variation of Sgr A$^\ast$, the time spans of the observations are 22:40UT - 03:40UT for the first day and 23:00UT - 05:45UT for the second day, providing good uv coverages for imaging. J1744-3116 was used as a phase calibrator. The flux density scale was determined using Titan and J1733-1304. 
Before spectral line imaging, the continuum emissions of Sgr A$^\ast$ and the GCMS were subtracted from the data using CASA task, {\tt mstransform (fitorder=1)} for each day of the observations.
Imaging for the combined data for the two-day observations  was done using CASA 5.0 with {\tt tclean} task to use a new auto-boxing capability.   
We used the multi-threshold auto-boxing algorithm ( {\tt usemask=`auto-multithresh'} ) to automatically identify the emission regions to be CLEANed using threshold based on rms noise and sidelobe level and update as the deconvolution iterations progress\footnote{https://casa.nrao.edu/casadocs/casa-5.0.0/synthesis-imaging/masks-for-deconvolution}.  For the weighting scheme, Briggs weighting with a robust parameter of 0.5 was used. As a result, we obtained an H30$\alpha$ recombination line data cube with high angular resolution ($0\farcs41 \times 0\farcs30, PA = -77^{\circ}$) and high sensitivity (0.2 mJy beam$^{-1}$ at a line-free channel).  The sensitivity is close to the expected one by the ALMA sensitivity calculator.

\begin{figure}
\figurenum{1}
\begin{center}

\includegraphics[width=15.5cm]{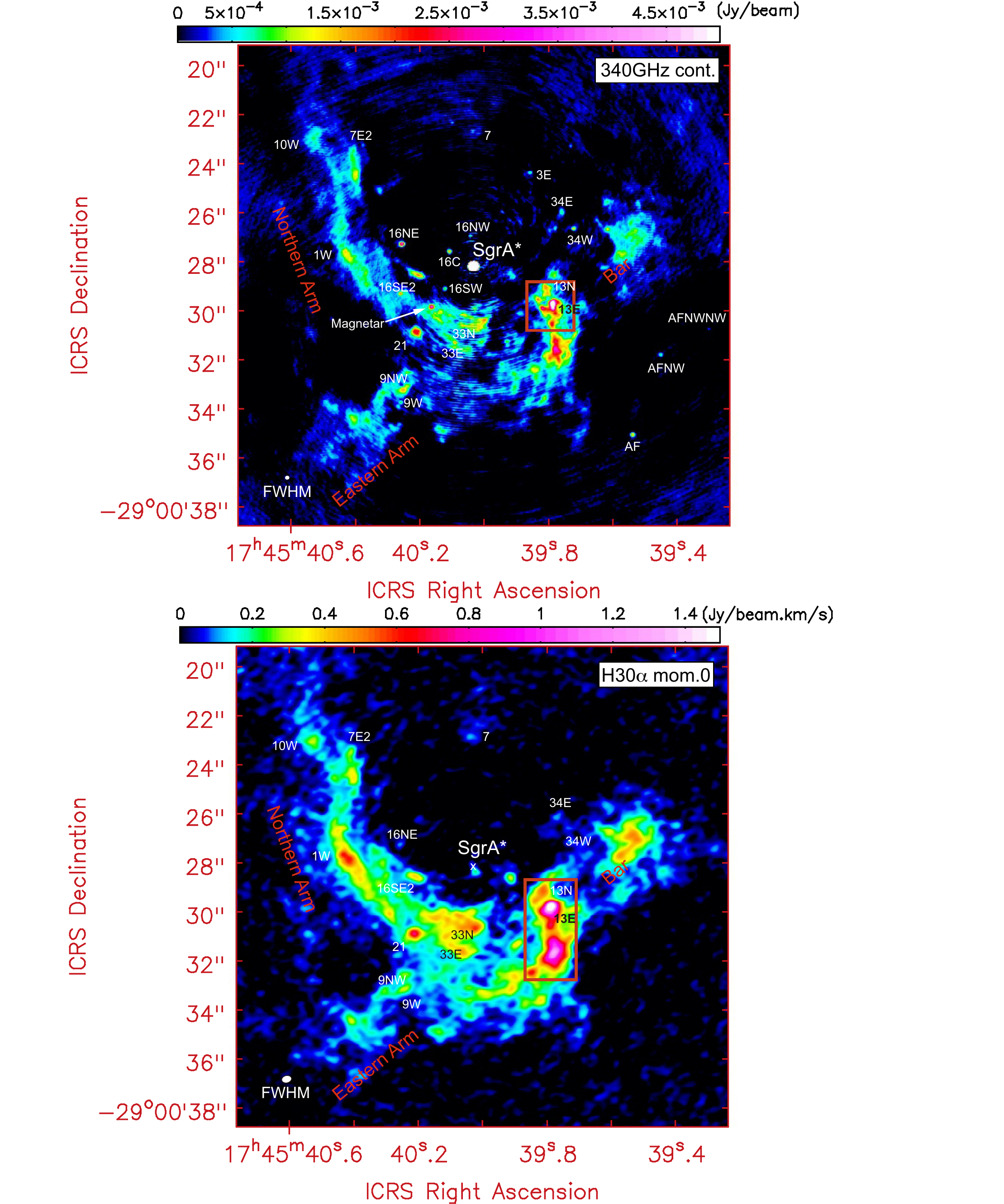}
\caption{The finding charts of the structures around  Sgr A$^\ast$ including the IRS13E complex.
{\bf Upper panel} Continuum map of Sgr A$^\ast$ and the GCMS at 340 GHz in units of Jy beam$^{-1}$.  The FWHM beam size is $0\farcs14 \times 0\farcs13, PA=82.5^\circ$ shown at the lower left corner.  
A red square indicates the area shown in Figure 2. 
{\bf Lower panel}  Integrated intensity (moment 0) map of the H30$\alpha$ recombination line with the integrated velocity range  from $V_{\mathrm {LSR}}= -400$ to $+400$ km s$^{-1}$ in units of Jy beam$^{-1}$ km s$^{-1}$. The FWHM beam size is $0\farcs41 \times 0\farcs30, PA=-77^\circ$ shown at the lower left corner. Sgr A$^\ast$ itself (cross) was not detected in the recombination line.  A red rectangle indicates the area shown in Figure 3. 
  }
\end{center}
\end{figure}
\begin{figure}
\begin{center}
\figurenum{2}
\includegraphics[width=15cm]{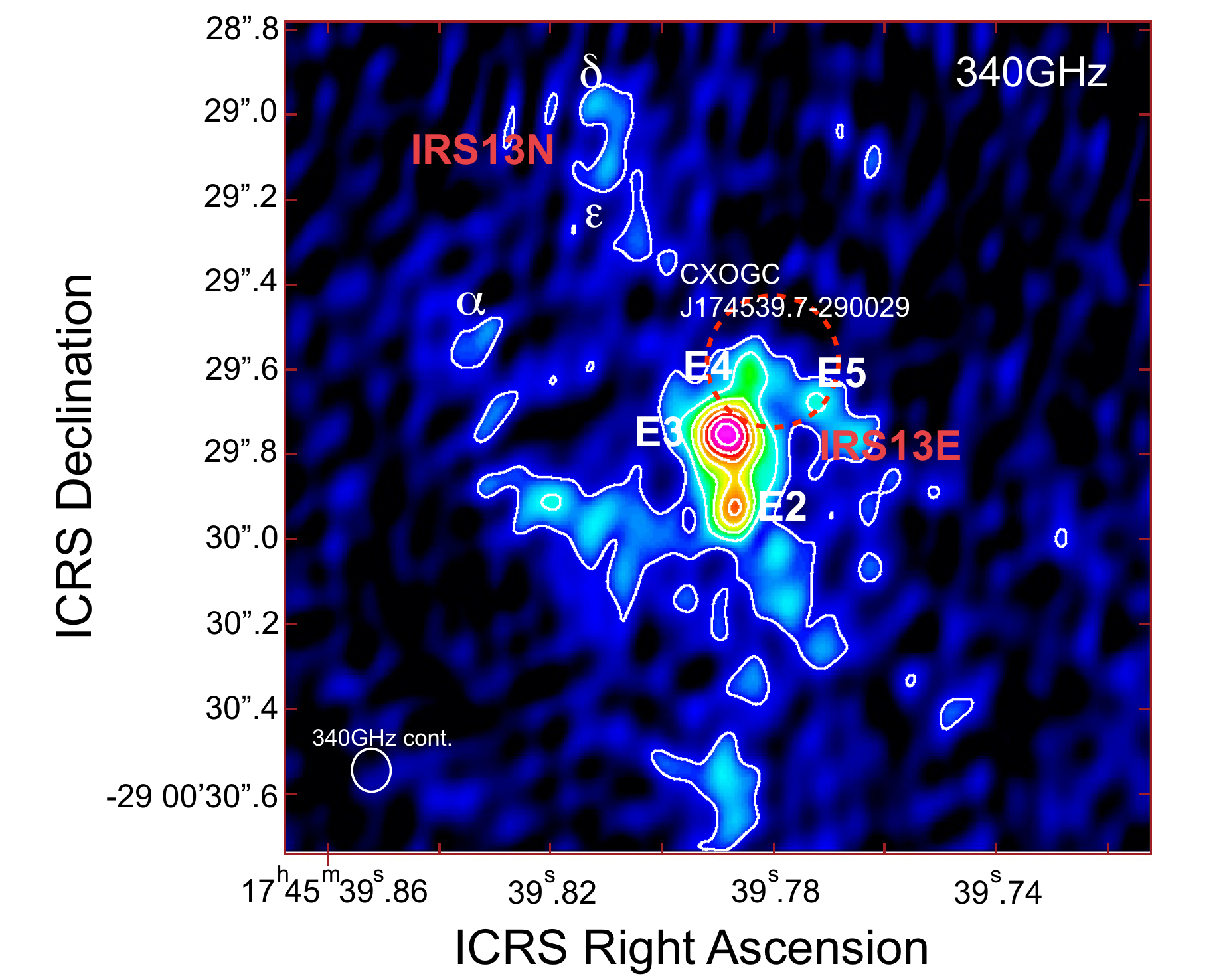}
\caption{340 GHz continuum emission map of the IRS13E complex with ALMA (red square in Figure 1-upper panel). The FWHM beam size is $0\farcs101 \times 0\farcs090, PA=3.8^\circ$, which is shown as the white ellipse at the lower left corner. The contour levels are 0.8, 1.6, 2.4, 3.2, 4.0, 4.8, and 5.6 mJy beam$^{-1}$ and the 1$\sigma$ noise level is 0.18 mJy beam$^{-1}$.
The red broken circle shows the position of the X-ray object: CXOGCJ174539.7-290029 ($\alpha_{\rm ICRS}=17^{\rm h}45^{\rm m}39\fs780$, $\delta_{\rm ICRS}=-29^\circ 00'29\farcs59$) \citep[e.g.][]{Baganoff, Muno}. The radius is the statistical error, $0\farcs16$. The IRS13N complex is also seen $0\farcs5$ northeast of the IRS13E complex. The Greek symbols show the members of the IRS13N complex.}
\end{center}
\end{figure}

\section{Results}
Figure 1 shows the continuum map at 340 GHz (upper panel) and the integrated intensity map of the H30$\alpha$ recombination line with the integrated velocity range  from -400 km s$^{-1}$ to +400 km s$^{-1}$ in $V_{\mathrm {LSR}}$ (lower panel).  They are the finding charts of the structures in the GCMS  \citep[Cf.][]{Tsuboi2016}. Although the flux density of Sgr A$^\ast$ was $S_\nu=2.8$ Jy at 340 GHz, Sgr A$^\ast$ itself was not detected in the recombination line. The IRS13E complex is the most prominent both in the maps. We concentrate on the IRS13E complex in this paper although we detected several fascinating objects in these maps, for example Magnetar PSR J1745-2900 \citep[e.g.][]{Eatough}  in the vicinity of Sgr A$^\ast$,  which is labeled ``Magnetar"  in the upper panel.  

Figure 2 shows the close-up  continuum map of the IRS13E complex at 340 GHz. 
The IRS13E complex is resolved into a group of compact objects in the map. Most of these are the IR identified objects \citep[e.g.][]{Maillard, Schodel2005, Paumard} including IRS 13E3 which is a main member of the  IRS13E complex but an ionized gas blob  as noted in Introduction \citep{Fritz2010}. 
The positions and angular sizes of these objects are estimated by  the two-dimensional Gaussian fit of CASA (see Table 1). 
 The relative position of IRS13E3 referring to Sgr A$^\ast$ is estimated to be $\Delta\alpha=-3\farcs19\pm0\farcs01, \Delta\delta=-1\farcs55\pm0\farcs01$.  The error is derived nominally according to ALMA Technical Handbook (10.6.6). 
The beam-deconvolved angular size of IRS13E3 is derived to be $\theta_{\mathrm{maj. obs.}}\times\theta_{\mathrm{min. obs.}}= (0\farcs102\pm0\farcs008)\times (0\farcs090\pm0\farcs008)$, $PA\sim27^\circ$, corresponding to the physical size of $0.0040$ pc $\times 0.0035$ pc ($800$ AU $\times 700$ AU) at the Galactic center distance.  The  estimated size is as small as the FWHM beam size. 
The total flux density of IRS13E3 is $S_{\nu {\mathrm 340}}=10.5\pm0.5$ mJy at 340 GHz by the Gaussian fit. Comparing with $S_{\nu {\mathrm 42}}=13.1$  mJy at 42 GHz \citep{Yusef-Zadeh2014}, the spectrum index is estimated to be $\alpha=log(S_{\nu {\mathrm 340}}/S_{\nu {\mathrm 42}})/log(340/42)\sim-0.1$. The flat spectrum suggests that the emission from IRS13E3 is an optically thin free-free emission and the sign of dust thermal emission is not clear.  
The derived parameters of the IR objects are also summarized in table 1.
Note that our derived values are similar to those in the previous observations \citep[e.g.][]{Yusef-Zadeh2014, Paumard}. 

\begin{table}
  \caption{Parameters of the compact objects in the IRS13E complex at 340 GHz. }
  \label{tab:first}
  \begin{center}
    \begin{tabular}{ccccccccc}
    \hline
Name&$\Delta\alpha^1$&$\Delta\delta^1$ &$\theta_{\rm maj.}\times \theta_{\rm min.}^{~2}$&$S_\nu({\rm peak})^3$& $S_\nu({\rm total})^3$ &Sp.$^4$\\
      &$[\arcsec]$&$[\arcsec]$& [mas$\times$mas, $PA$] & [mJy beam$^{-1}$] &  [mJy beam$^{-1}$] &\\
\hline
Sgr A$^\ast$&$0.00$ &$0.00$& point source &$2807\pm2^5$&$2811\pm3^5$&-- \\

IRS13E2&$-3.21\pm0.02^6$ &$-1.70\pm0.02$& point source &$3.1\pm0.3$&$5.0\pm0.6$&WN9 \\

IRS13E3&$-3.19\pm0.01$ &$-1.55\pm0.01$& $102\times 90, 27^\circ$&$5.4\pm0.2$&$10.5\pm0.5$&Ionized Gas \\

IRS13E4&$-3.24\pm0.02$ &$-1.42\pm0.02$& $97\times 77,176^\circ $&$2.0\pm0.2$&$3.6\pm0.5$&WC9 \\

IRS13E5&$-3.40\pm0.04$ &$-1.47\pm0.04$& point source &$1.1\pm0.1$&$1.1\pm0.2$ &Ionized Gas \\

\hline
    \end{tabular}
 \end{center}
 $^1$ Peak position referring to Sgr A$^\ast$ estimated by the two-dimensional Gaussian fit of CASA.
          The position of Sgr A$^\ast$ is assumed to be $\alpha_{\rm ICRS}$ = $17^{\rm h}45^{\rm m}40\fs04$ and $\delta_{\rm ICRS}$= $-29^{\circ}00'28\farcs20$.
$^2$ Beam-deconvolved size.
 $^3$ Flux estimated by the two-dimensional Gaussian fit of CASA.
 $^4$ IR spectral type \citep{Fritz2010}.  
$^5$ Because Sgr A$^\ast$ is a variable source, these are averages at Aug. and Sep. 2016. 
$^6$ The error is derived nominally according to ALMA Technical Handbook (10.6.6).
\\
\clearpage
\end{table}
\begin{figure}
\figurenum{3}
\begin{center}
\includegraphics[width=18cm]{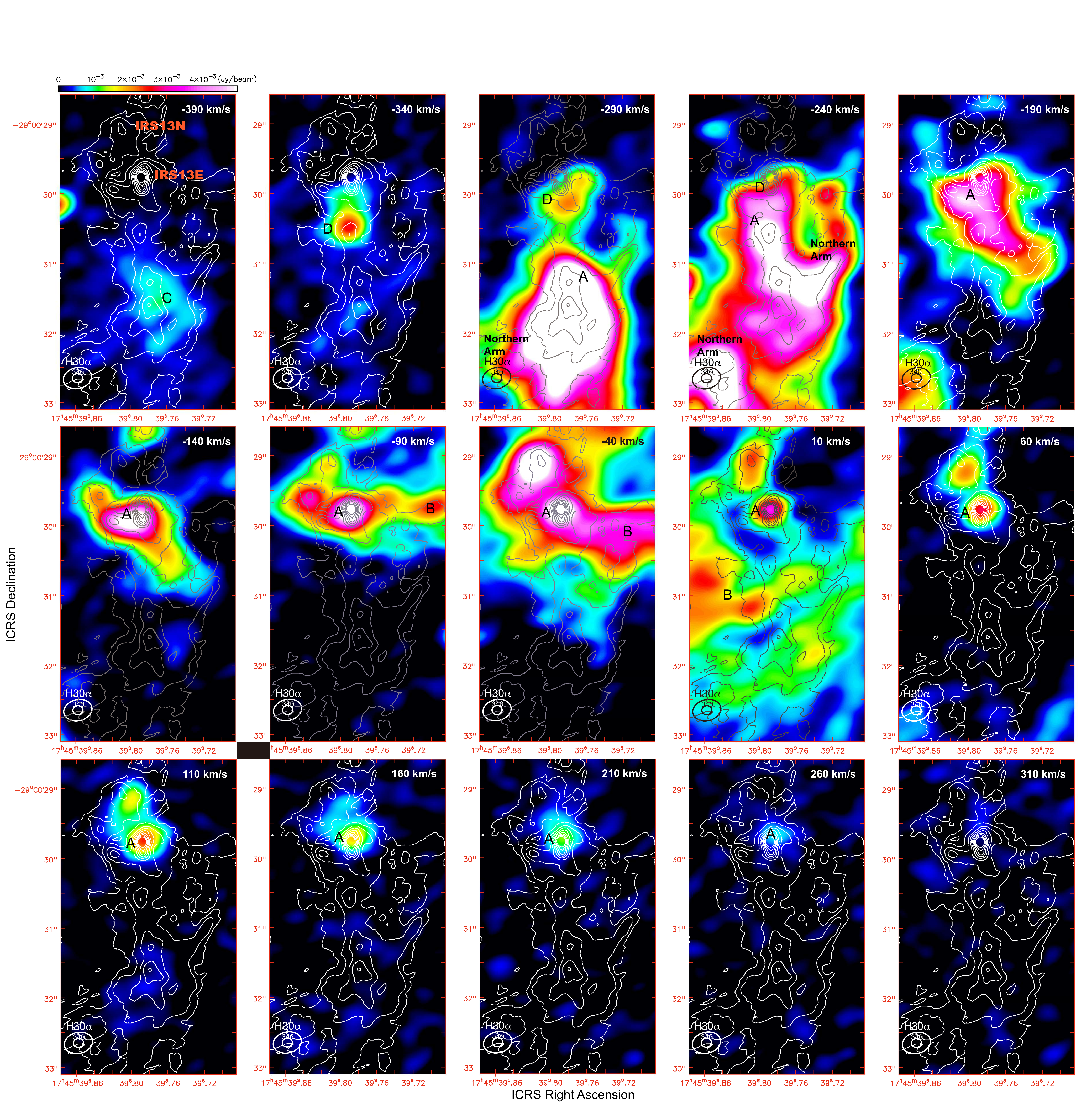}
\caption{Channel maps of the IRS13E complex in the H30$\alpha$ recombination line from -415 km s$^{-1}$ to $+335$ km s$^{-1}$ in $V_{\mathrm {LSR}}$ (red rectangle in Figure 1-lower panel). The velocity width of each panel is 50 km s$^{-1}$. The central velocity and the FWHM beam size, $0\farcs41 \times 0\farcs30, PA=-74^\circ$, are shown at the upper right and lower left corners of each panel, respectively. 
The 340 GHz continuum emission is also shown as contours for comparison.  The FWHM beam size, $0\farcs14 \times 0\farcs13, PA=82.5^\circ$, at 340 GHz is also shown at the lower left corner of each panel. The contour levels are 0.4, 0.8, 1.6, 2.4, 3.2, 4.0, 4.8, 5.6, and 6.4 mJy beam$^{-1}$. The capitals show the components shown in figure 4. }
\end{center}
\end{figure}

Figure 3 shows the channel maps of the IRS13E complex  in the H30$\alpha$ recombination line with the central velocities of $V_{\mathrm {c, LSR}}=-390$ to $+310$ km s$^{-1}$ (the map area corresponds to the red rectangle in Figure 1-lower panel). The velocity width of the channel maps is $\Delta v=50$ km s$^{-1}$.  Figure 4a shows the integrated intensity map of the H30$\alpha$ recombination line with the integrated velocity range of  $V_{ \mathrm {LSR}}=-400$ to $+400$ km s$^{-1}$. These maps also show the 340 GHz continuum emission as the contours. The ionized gas seen in the panels with $V_{ \mathrm {c, LSR}}\sim -340$ to $+260$ km s$^{-1}$ seems to be along a continuum north-south ridge including the IRS13E complex. Figure 4b shows the position velocity diagram of the H30$\alpha$ recombination line through the IRS13E complex along the red rectangle shown in figure 4a. 

There is a compact component located at $\sim1"$ south of IRS13E3 in the panel with $V_{\mathrm {c, LSR}}=-340$  km s$^{-1}$ of Figure 3,  which is labeled ``D".  The compact component is approaching to IRS13E3 with increasing velocity (see the panels with $V_{\mathrm {c, LSR}}=-340, -290,$ and $-240$  km s$^{-1}$).  This is also identified as a component in the position velocity diagram. 
An extended component appears around $\sim2\arcsec$ south of IRS13E3 in the panel with $V_{\mathrm {c, LSR}}=-290$  km s$^{-1}$, which is labeled ``A".
The component also shifts to north with increasing velocity and reaches to IRS13E3 in the panel with $V_{ \mathrm {c, LSR}}\sim -190$ km s$^{-1}$.  The components D and A seem to be combined at IRS13E3 and the combined compact component stays here in the velocity range of $V_{\mathrm {c, LSR}}=-90$ to $+260$ km s$^{-1}$.
The peak position seems to shift monotonically to the north across the IRS13E3 in the velocity range of $-90$ to $+210$ km s$^{-1}$. The peak shift is shown clearly in Figure 4d.
The behavior is also identified as a prominent curved ridge with a wide velocity width in Figure 4b. 
The negative angular offset part of the curved ridge corresponds to the south extended component in the panels with $V_{ \mathrm {c, LSR}}\sim -290$ to $-190$ km s$^{-1}$ of Figure 3. 
The velocity width of the ionized gas toward IRS13E3 reaches to $\Delta v_{\mathrm{FWZI}} \sim 650$ km s$^{-1}$ (Full Width at Zero Intensity; FWZI) at  IRS13E3.

\begin{figure}
\begin{center}
\figurenum{4}
\includegraphics[width=16cm]{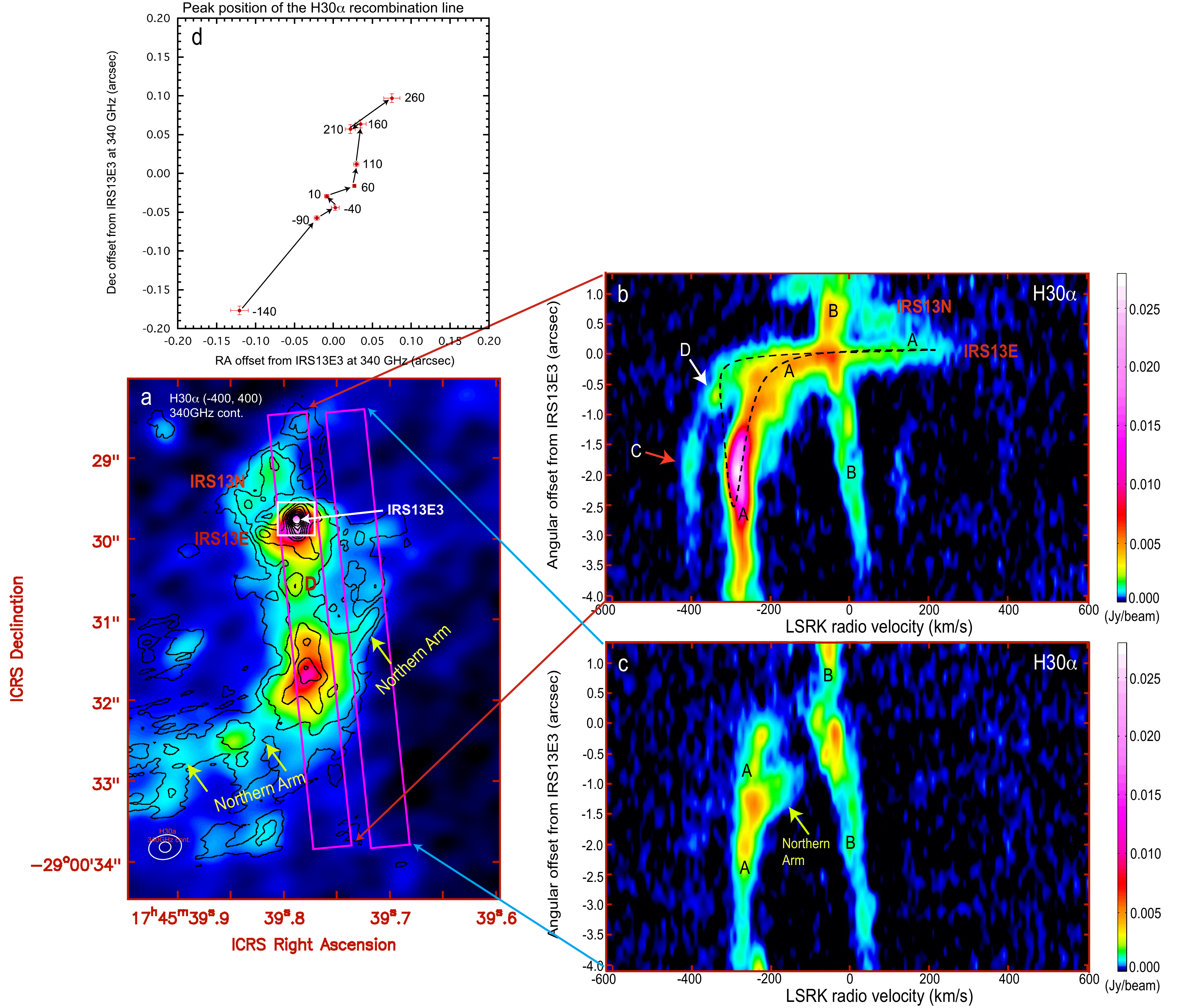}
\caption{
{\bf a} Integrated intensity map of the H30$\alpha$ recombination line with the integrated velocity range of  $V_{ \mathrm {LSR}}=-400$ to $+400$ km s$^{-1}$.  The contours show the 340 GHz continuum emission with ALMA. The FWHM beam sizes are shown as white ovals at the lower left corner.  The red rectangle shows the cutting area of the position-velocity diagram of panel b. The white square shows the area of panel c. Yellow arrows show the tip of the Northern Arm (NA) of the Minispirals. 
{\bf b} Position velocity diagram of the H30$\alpha$ recombination line through the IRS13E complex along the cutting area shown in the panel a in units of Jy beam$^{-1}$. 
There is a curved ridge with a wide velocity width (``A"), which reaches to $\Delta v_{\mathrm{FWZI}} \sim 650$ km s$^{-1}$ (Full Width at Zero Intensity; FWZI) at  IRS13E3,  
and a relatively narrow velocity width component around $V_{\mathrm {LSR}}\sim-50$ to $+50 $km s$^{-1}$ (``B"). A black broken line shows an example of the Keplerian orbits that describe well the observed features.
The eccentricity, the semi-major axis, and the position angle  are $e\sim0.97$, $a\sim1.4\times10^{17}$ cm, and $PA\sim60^\circ$, respectively. {\bf c} Position velocity diagram of the H30$\alpha$ recombination line adjacent parallel to {\bf b}. Yellow arrow shows the tip of the NA. 
{\bf d} Peak positions of of the H30$\alpha$ recombination line derived by 2-D Gaussian fit.  Numbers show the $V_{\mathrm {c, LSR}}$.  }
\end{center}
\end{figure}

A nearly vertical ridge with $V_{\mathrm {LSR}}\sim-50$ to $+50$ km s$^{-1}$ is identified in the position velocity diagram of Figure 4b,  which is labeled ``B".  This  corresponds to an extended component surrounding the IRS13E and IRS13N complexes seen in the channel map of $V_{ \mathrm {c, LSR}}=-90$ to $10$ km s$^{-1}$ of Figure 3. The component is a part of the Bar which is a substructure of the GCMS (also see Figure 1). 
The ionized gas probably associated with the IRS13N complex is also identified in the panels with $V_{ \mathrm {c, LSR}}\sim-90$ to $+110$ km s$^{-1}$ of Figure 3.

In addition, a faint ridge is seen in the position velocity diagram of Figure 4b, which is labeled ``C". The component is the counterpart in the He30$\alpha$ recombination line of the negative velocity vertical part of the components ``A" because the velocity shift is exactly equal to the velocity difference between the H30$\alpha$ and He30$\alpha$ recombination lines; $\Delta v=c[\nu(\mathrm {He}30\alpha)-\nu(\mathrm {H}30\alpha)]/\nu(\mathrm {H}30\alpha)\sim -122$  km s$^{-1}$. This component is also identified as a faint extended component $\sim2\arcsec$ south of IRS13E3 in the channel map of $V_{ \mathrm {c, LSR}}\sim -390$ km s$^{-1}$(see Figure 3).
The intensity ratio of $\frac{I(\mathrm {He}30\alpha)}{I(\mathrm {H}30\alpha)}$ has a reasonable value of $\sim0.05$  in galactic disk HII regions \citep[e.g.][]{Roshi}.

Figure 4c shows the position velocity diagram of the H30$\alpha$ recombination line along the cut adjacent and parallel to the cut for Figure 4b for comparison. This indicates the relations between the ionized gas associating with the IRS13E complex and the arms of the GCMS. The tip of the Eastern Arm (EA) is located around $\alpha_{\rm ICRS}=17^{\rm h}45^{\rm m}40\fs03$, $\delta_{\rm ICRS}=-29^\circ 00'30\farcs5$ (see the lower panel of Figure 1), which is far from the ionized gas associating with the IRS13E complex. Therefore, there is no connection with the EA and the ionized gas. The ionized gas is independent from the EA.  On the other hand, the relation between the ionized gas and the Northern Arm (NA) is complicated. The tip of the NA crosses the ionized gas around $2"$ south of IRS13E as shown in Figure 4a (yellow arrows, see also  the panels with $V_{ \mathrm {c, LSR}}\sim -290$ to $-190$ km s$^{-1}$ of Figure 3). Although the tip of the NA is partially blended with the ionized gas, this is barely identified in the position velocity diagram of Figure 4c (yellow arrow). The ionized gas would be independent from the NA. 

\section{Discussion}

\subsection{Compactness of the Ionized Gas toward  IRS13E3}

The angular source size of the ionized gas toward  IRS13E3 is derived to be $\theta_{\mathrm{maj. obs.}}\times\theta_{\mathrm{min. obs.}}= 0\farcs41\pm0\farcs03\times 0\farcs31\pm0\farcs02$, $PA\sim127^\circ$  by the two-dimensional Gaussian fit to the total integrated velocity map (see Figure 1-lower panel). 
Because this is as large as the clean beam size of the H30$\alpha$ recombination line, the beam-deconvoled source size is estimated to be much less than the beam size.
The peak position is also estimated to be $\alpha_{\rm ICRS}=17^{\rm h}45^{\rm m}39\fs785\pm0\fs001$, $\delta_{\rm ICRS}=-29^\circ 00'29\farcs84\pm0.\arcsec01$.  This position is almost equal to that of IRS13E3 in the 340 GHz continuum map within the FWHM beam size.  The positional differences are $\Delta \alpha\sim0\farcs005$ and $\Delta \delta\sim0\farcs028$ according to the comparison between the 340 GHz continuum and H30$\alpha$ moment 0 positions of IRS16NE, of which images are compact both in the maps (see Figure 1). The positional correspondence between both the observation is better than 10\% of the FWHM beam size of  the H30$\alpha$ moment 0 map ($\sim0\farcs4$). 

Because IRS13E3 is emitting the 340 GHz continuum through thin free-free emission mechanism as mentioned above, IRS13E3 is emitting the  H30$\alpha$ recombination line simultaneously. 
The area emitting the recombination line should be identical to IRS13E3 itself shown in the continuum map (see Figure 2).   The size of the ionized gas would be $\theta_{\mathrm{maj. obs.}}\times\theta_{\mathrm{min. obs.}}= 0\farcs102\times 0\farcs090$, $PA\sim27^\circ$ or $r_{\mathrm{maj}}\times r_{\mathrm{min. obs.}}= 0.0020$ pc $\times 0.0018$ pc  ($400$ AU $\times 350$ AU). Although our estimated radius, $r$, is an upper limit, this radius is considered to be close to the real radius because  the continuum observation of IRS13E3 with JVLA at 34 GHz  shows a similar source size; $0\farcs08 \times 0\farcs04$ \citep{Yusef-Zadeh2014}.  
The compactness and large velocity width suggest the presence of an IMBH in the IRS13E complex.

\subsection{Keplerian Orbit with High Eccentricity around IRS13E3?}

There is a bright ridge connecting IRS13E3 with the extended component around $\sim 2\arcsec$ south of IRS13E3  in Figure 4a. 
This corresponds to the curved ridge from $V_{ \mathrm {LSR}}\sim-300$ to $+250$ km s$^{-1}$ (``A") in the position velocity diagram of figure 4b. 
In addition, there is also a weak component at $V_{ \mathrm {LSR}}\sim -350$ km s$^{-1}$ at the offset of $\sim -0\farcs5$ (``D"; white arrow) in the position velocity diagram. 
The weak component is identified as a compact component at $0\farcs8$ south IRS13E3 in the channel map of $V_{ \mathrm {c, LSR}}= -340$  km s$^{-1}$ as mentioned in the previous section (see Figure 3). 
These observed features suggest the presence of a Keplerian orbit with high eccentricity around IRS13E3.
Such Keplerian orbits have a nearly linear part with large velocity gradient and double curved ridges with relatively small velocity gradient in the position velocity diagram \citep[see Figure 12 in][]{Tsuboi2017}. 
In the case, the line intensity on the orbit decreases with increasing velocity as shown in Figure 4b because the line intensity should be in inverse proportion to the orbital velocity. 
Note that the thermal velocity broadening for ionized Hydrogen gas of $10^4$ K is $\Delta v\sim 20$  km s$^{-1}$ in FWHM, which is negligible as compared with the rotation velocity. 

Another hypothesis to explain the weak component D may be that the component is the counterpart in the He30$\alpha$ recombination line.
However, the intensity ratio of the bright ridge and weak component is up to $\sim0.3$. This is too high as $\frac{I(\mathrm {He}30\alpha)}{I(\mathrm {H}30\alpha)}$. In addition, the velocity shift between these components is $\Delta v \sim-100$ km s$^{-1}$ which is less than the velocity difference between the H30$\alpha$ and He30$\alpha$ recombination lines, $\Delta v \sim-122$ km s$^{-1}$, as mentioned previously. Therefore, the weak component would be a part of the Keplerian orbit with a high eccentricity around  IRS13E3. 

An example of the orbits  describing well the observed features is shown in the position velocity diagram (see the black broken line in Figure 4b). 
Note that it is difficult to determine exclusively the accurate orbital parameters by fitting in the position velocity diagram because the full orbit is not completely occupied by ionized gas and/or the observed features are not always belong to a single orbit.
The apoastron of the orbit would be located in the bright extended ionized gas component seen $\sim2\arcsec$  south of the IRS13E complex in projection. The angle between the direction of the semi-major axis and the line of sight is $PA\sim60^\circ$. The observed major axis is $2a\sin60^\circ\sim2.5\times10^{17}$ cm. Then the semi-major axis of the orbit is estimated to be $a\sim1.4\times10^{17}$ cm $=1\times10^4$ AU.
The inclination angle of the orbit would be $i\sim0^\circ$ because of the elongated feature with a narrow width of the orbiting ionized gas shown in Figure 4a. By the comparison between the observed shape in the position velocity diagram and  the calculated Keplerian orbits, the eccentricity of the orbit is estimated to be $e\sim0.97$.

\subsection{The Enclosed Mass of IRS13E3}
 The presence of an IMBH in the IRS13E complex is strongly supported by a large enclosed mass of IRS13E3. In the case of nearly edge-on view, the velocity width of the Keplerian orbit is estimated by 
\begin{equation}
\label{1}
\Delta V \sim \Big(\sqrt{\frac{1+e}{1-e}}+\sqrt{\frac{1-e}{1+e}} \Big)\sqrt{\frac{GM}{a}}.
\end{equation}

Then the enclosed mass is estimated to be
\begin{equation}
\label{2}
M \sim a\Delta V^2 \Big(\sqrt{\frac{1+e}{1-e}}+\sqrt{\frac{1-e}{1+e}} \Big)^{-2}G^{-1}.
\end{equation}
The observed velocity width is $\Delta V\sim 600$ km s$^{-1}$, and the enclosed mass of the object is estimated to be $M\sim4-7\times10^{4}$ M$_\odot$, comparable to the mass of an IMBH,  for $e=0.96-0.98$ and $a\sim1.4\times10^{17}$ cm.  
Although the Keplerian orbit has large ambiguities in the orbit parameters, the mass would be consistent with $M\sim10^{4}$ M$_\odot$ in the IRS13E complex \citep[e.g.][]{Maillard, Schodel2005, Paumard}.  
Even in the case that the compact ionized gas is not belong to the Keplerian orbit, the compactness and large velocity width would indicate $\sim10^{4}$ M$_\odot$ in the IRS13E complex.

If there is an IMBH orbiting around Sgr A$^\ast$, the position of Sgr A$^\ast$ must be  affected by it. The position of Sgr A$^\ast$ on the celestial sphere has been monitored  using VLBA for a long time over 15 years \citep[e.g.][] {Reid}. The positional shift along the Galactic plane is as large as -6.4 mas year$^{-1}$. However, one can not distinguish between the perturbation by the massive object and the proper motion by the Galactic rotation. On the other hand, the positional shift crossing the Galactic plane is found to be as small as -0.2 mas year$^{-1}$.  This indicates that the upper limit mass of the second black hole is $M\lesssim 10^{4}$ M$_\odot$ in the area of $r\sim 10^3-10^5$ AU from Sgr A$^\ast$ \citep{Reid}. Therefore, there is no conflict between our derived enclosed mass and the upper limit mass of the second black hole inferred from the VLBA observations. 

The periastron distance is estimated to be $a(1-e)=4.3\times10^{15}$ cm and the projected distance from Sgr A$^\ast$ is $r_{\mathrm{proj.}}\sim4\times10^{17}$cm. Because the mass of the IMBH is only $1$\% of the mass of Sgr A$^\ast$, the gravity of  the IMBH associated with IRS13E3 is comparable to that of  Sgr A$^\ast$ even around the periastron of the Keplerian orbit around IRS13E3. However, the flow of the ionized gas, which is extended to south, seems to be described by a Keplerian orbit as mentioned above.  This means that the gravity of the IMBH would be dominant there, and the IMBH might be located fairy far from Sgr A$^\ast$ than the projected distance.

The ionized gas mass on the orbit is estimated to be $M\sim6\times10^{-3}$M$_\odot$ based on this observation assuming the electron temperature of $T_{\mathrm e}\sim1\times10^4$ K and the electron density of $n_{\mathrm e}\sim1\times10^6$ cm$^{-3}$ \citep[e.g.][]{Murchikova, Tsuboi2017}. When all the ionized gas falls to the IMBH associated with IRS13E3 within one orbital period; $T=2\pi\sqrt{\frac{a^3}{GM}}\sim4\times10^3$ yr, the upper limit of the mass accretion rate is estimated to be $M/T\lesssim1\times10^{-6} $~M$_\odot$ yr$^{-1}$. However, because both approaching gas to IRS13E3 (probably  D)  and going-away gas (probably  A) are observed on the orbit \citep[Cf.][]{Oka}, it would be overestimation that all gas falls into it within an orbital period. 
If  the efficiency of the accretion is assumed to be 1 \% as a likely value, the mass accretion rate  becomes  comparable to the mass accretion rate of Sgr A$^\ast$ \citep[e.g.][]{Quataert, Genzel2010}.

\subsection{X-ray Counterpart of IRS13E3?}
Figure 2 also shows the position of a discrete X-ray source detected in the IRS 13E complex, CXOGCJ174539.7-290029 ($\alpha_{\rm ICRS}=17^{\rm h}45^{\rm m}39\fs780$, $\delta_{\rm ICRS}=-29^\circ 00'29\farcs59$) \citep[e.g.][]{Baganoff, Muno}. The statistical error and absolute  uncertainty of the position are $0\farcs16$ and $0\farcs76$ in radius, respectively \citep[][]{Baganoff}.
The X-ray source is located near the 340-GHz continuum peak position of IRS13E3 with a positional difference of $\sim0\farcs2$  (see Table 1), which is smaller than the uncertainty of the X-ray observation.
Such a small positional difference is generally caused by using different reference frames. 
Therefore, we consider that the two positions are coincident.
 In addition, in the case that the compact component is a part of the ionized gas orbiting around the periastron as shown previously, it is possible that the component and the X-ray source corresponding to the IMBH are observed to be up to  the periastron distance apart ($\sim0\farcs3$). 

The X-ray photon flux is $F_\mathrm{2-8 keV}=3.092\times10^{-7}$  cm$^{-2}$ s$^{-1}$  at 1999.72 yr, which is a half of that of Sgr A$^\ast$ \citep[e.g.][]{Muno}.  The spectrum of this bright X-ray source resembles that of the quiescent emission from the hot plasma around Sgr A$^\ast$ and the source has no long- and short-timescale variabilities \citep[][]{Muno}.  The X-ray emission should be originated by the hot plasma located at IRS13E3 through bremsstrahlung  rather than those from usual X-ray binaries, cataclysmic variables, and so on.  Thus we consider that this source is the X-ray counterpart of  IRS13E3.   The large X-ray photon flux may be consistent with the mass accretion rate estimated in the previous subsection.

\section{Conclusions}
We detected the ionized gas associated with IRS13E3 in the H30$\alpha$ recombination line using ALMA, which has a wide velocity width   ($\Delta v_{\mathrm{FWZI}} \sim 650$ km s$^{-1}$) and compactness ($r\sim0.002$ pc=$400$ AU).  The enclosed mass is estimated to be $10^{4}$ M$_\sun$ in the case of the  Keplerian orbit with high eccentricity around IRS13E3. 
The mass does not conflict with the upper limit mass of the IMBH around Sgr A$^\ast$ which is derived by the long-term astrometry with VLBA. This object presumably has an X-ray counterpart. 
Consequently, a very fascinated possibility is that the detected ionized gas is orbiting around an IMBH with $10^{4}$ M$_\odot$ embedded in the IRS13E complex.
 
 \acknowledgments
This work is supported in part by the Grant-in-Aid from the Ministry of Eduction, Sports, Science and Technology (MEXT) of Japan, No.16K05308. The National Radio Astronomy Observatory is a facility of the National Science Foundation operated under cooperative agreement by Associated Universities, Inc. USA. This paper makes use of the following ALMA data:ADS/JAO.ALMA\#2015.1.01080.S and ALMA\#2015.A.00021.S. ALMA is a partnership of ESO (representing its member states), NSF (USA) and NINS (Japan), together with NRC (Canada), NSC and ASIAA (Taiwan), and KASI (Republic of Korea), in cooperation with the Republic of Chile. The Joint ALMA Observatory is operated by ESO, AUI/NRAO and NAOJ. 

\vspace{5mm} 
\facilities{ALMA}
\software{CASA}

\clearpage

\end{document}